\documentclass[aps,prx,amsmath,floats,floatfix,twocolumn,
  superscriptaddress,nofootinbib,showpacs]{revtex4-1}

\pdfoutput=1 
\usepackage{diagbox}
\usepackage{mathrsfs}
\usepackage{multirow}
\usepackage{braket}
\usepackage{amsfonts}
\usepackage{amsmath}
\usepackage{amssymb}
\usepackage{amsthm}
\usepackage{bm}
\usepackage{graphicx}
\usepackage{graphics}
\usepackage{latexsym}
\usepackage{rotating}
\usepackage[dvipsnames]{xcolor}
\usepackage{mathrsfs}
\usepackage{microtype}
\usepackage{verbatim}
\usepackage{url}

\usepackage[caption=false]{subfig}
\usepackage[normalem]{ulem}

\usepackage[breaklinks=true]{hyperref}
\hypersetup{
    colorlinks=true,
    linkcolor=NavyBlue,
    filecolor=Magenta,
    urlcolor=NavyBlue,
    citecolor=NavyBlue
}

\usepackage{yfonts}
\usepackage{float}
\usepackage{xspace}
\usepackage{mathrsfs}
\usepackage{siunitx}
\usepackage{array}

\allowdisplaybreaks[4]

\makeatletter
\newcommand*{\rom}[1]{\expandafter\@slowromancap\romannumeral #1@}
\makeatother

\AtBeginDocument{%
    \newwrite\bibnotes
    \def\bibnotesext{Notes.bib}
    \immediate\openout\bibnotes=\jobname\bibnotesext
    \immediate\write\bibnotes{@CONTROL{REVTEX41Control}}
    \immediate\write\bibnotes{@CONTROL{%
    apsrev41Control,author="08",editor="1",pages="1",title="0",year="1"}}
    \if@filesw
    \immediate\write\@auxout{\string\citation{apsrev41Control}}%
    \fi
}

\allowdisplaybreaks

\usepackage{newtxtext,newtxmath}

\begin{document}

\title{Spinning Black Holes in Modified Gravity via Spectral Methods}

\author{Kelvin~Ka-Ho~Lam} 
\email{khlam4@illinois.edu}
\affiliation{Illinois Center for Advanced Studies of the Universe \& Department of Physics, University of Illinois Urbana-Champaign, Urbana, Illinois 61801, USA}

\author{Adrian Ka-Wai Chung}
\email{akwchung@illinois.edu}
\email{kwc43@cam.ac.uk}
\affiliation{Illinois Center for Advanced Studies of the Universe \& Department of Physics, University of Illinois Urbana-Champaign, Urbana, Illinois 61801, USA}
\affiliation{DAMTP, Centre for Mathematical Sciences, University of Cambridge, Wilberforce Road, Cambridge CB3 0WA, United Kingdom}

\author{Nicol\'as Yunes}
\affiliation{Illinois Center for Advanced Studies of the Universe \& Department of Physics, University of Illinois Urbana-Champaign, Urbana, Illinois 61801, USA}

\date{\today}

\begin{abstract} 


Rapidly-rotating black-hole spacetimes outside general relativity are key to many tests of Einstein's theory.
We here develop an efficient spectral method to represent such spacetimes analytically, in closed-form, and to high accuracy, in a large class of effective-field-theory extensions of general relativity.  
We exemplify this method by constructing, for the first time, closed-form and analytic representations of spinning black holes in scalar-Gauss-Bonnet, dynamical Chern-Simons, and axidilaton gravity to an accuracy better than $10^{-8}$ for all dimensionless spins below 0.99.

\end{abstract}

\maketitle

\vspace{0.2cm}
\noindent 
\textit{Introduction.}\textemdash \
Einstein's theory of general relativity is among the most successful theoretical frameworks of modern physics, passing many stringent tests in the Solar System and in binary pulsars for more than a century~\cite{Will2014, Yunes:2013dva}. 
Despite these successes, general relativity is not a complete description of nature: it is incompatible with quantum theory at the smallest scales~\cite{tHooft:1974toh, Deser:1974cz, Deser:1974xq}, it cannot explain black-hole singularities~\cite{Hawking:1970zqf}, and it cannot naturally account for cosmic inflation in the early Universe~\cite{Guth:1980zm, Starobinsky:1980te}, the late-time acceleration of the universe without invoking dark energy or an unnaturally-small cosmological constant~\cite{Weinberg:1988cp}, or galactic rotation curves without dark matter~\cite{Sofue:2000jx, Rubin:1970zza, Bertone:2018krk}.
Such limitations have motivated the study of extensions to Einstein’s theory, many of which can now be tested with the rapidly-growing catalog of gravitational-wave observations.
These data provide a unique opportunity to probe the ``extreme gravity regime,'' where the spacetime curvature is simultaneously strong and dynamical, such as during black-hole mergers and neutron-star collisions.

Thanks to the LIGO-Virgo-KAGRA collaboration and the Event Horizon Telescope, we can now probe the extreme-gravity dynamics of compact objects through observations of gravitational wave and electromagnetic events. 
Gravitational waves emitted by binary black hole mergers are generated in the vicinity of their event horizons, and electromagnetic emissions from accretion disks probe regions close to the innermost stable circular orbit of supermassive black holes \cite{Abramowicz:2011xu}. 
These signals grant us a magnifying glass with which to test general relativity in the high-curvature regime of astrophysical black holes and to search for deviations predicted by modified gravity at those scales. 
However, computing theory-specific observables remains a major challenge. 
Such theories typically involve higher-derivative corrections to the Einstein equations \cite{Bueno:2016xff, Burger:2019wkq, Dehghani:2011vu} and may include additional dynamical scalar or vector fields \cite{Cano_Ruiperez_2019, Kanti:1995cp, Jackiw:2003pm}, leading to a system of coupled, nonlinear, elliptic partial differential equations. 
The absence of exact black hole solutions in these theories significantly hinders our ability to confront them with observational data.

Significant efforts have been put forth to construct rotating black holes in modified gravity theories, inspired by effective-field theory. 
For example, numerical solutions were computed in scalar Gauss-Bonnet gravity \cite{Delgado:2020rev, Kleihaus:2011tg, Kleihaus:2015aje}, dynamical Chern-Simons gravity \cite{Sullivan:2019vyi, Sullivan:2020zpf}, cubic gravity \cite{Burger:2019wkq}, and quartic gravity \cite{Dehghani:2011vu} using finite difference methods. 
Due to their nature, however, these numerically-approximate solutions are not expressed in terms of analytic, closed-form functions, requiring a large number of spatial grid points to achieve sufficient precision for smooth interpolation (especially when in need for their derivatives).
Analytic, closed-form, approximate solutions have only been found using the small-coupling and slow-rotation expansion, which enable a perturbative treatment in both the coupling constant and spin parameter $a$ \cite{Yunes:2009hc, Yagi:2012ya, Maselli:2015tta, Cano_Ruiperez_2019, Cardoso:2018ptl, Cano:2021myl}. 
This scheme greatly reduces the complexity of the modified field equations, thus allowing for closed-form solutions, but many astrophysical black holes are observed to have moderate or large spins, rendering ``series-in-$a$ solutions'' insufficient \cite{KAGRA:2021vkt, LIGOScientific:2025slb, Reynolds:2013rva}. 
While, in principle, this problem could be eliminated by extending the spin expansion to a higher order (assuming the series-in-$a$ expansion is absolutely convergent), the computational cost grows rapidly, making such an approach impractical. 

To overcome all of these limitations, we here develop a new framework, based on spectral and pseudospectral (collocation) methods, to construct analytic and closed-form representations of highly-accurate, approximate black hole solutions with any spin in modified gravity \cite{Fernandes:2022gde, Liu:2025mfn}. 
Our spectral framework consists of first representing the spacetime as a linear, beyond-Einstein deformation of a Kerr black hole\footnote{EFTs require solutions to be a smooth deformation of GR solutions. This implies that nonperturbative solutions, such as scalarized black-holes, could not be constructed with our methodology.}, and then peeling off certain analytic behavior from the deformations to ensure certain boundary conditions are satisfied. 
The metric deformations are then represented as spectral expansions, which allow the field equations to be decomposed into a set of algebraic equations. 
For a given spin value, these algebraic equations can be solved by standard matrix techniques to obtain the spectral expansion coefficients. 
Repeating this approach for many spin values, we obtain the spectral expansion coefficients as a discrete function of spin, which can then be represented as a polynomial in the logarithm of the extremality parameter $\delta \equiv 1 - a^2$. 
Such a framework \emph{never} requires an expansion in small spin or the numerical evaluation of spectral coefficients at every spacetime grid point, instead providing an analytic, closed-form expression (in a given Boyer-Lindquist-like coordinate system) for every component of the metric that is accurate for any spin value \cite{Lam:2025fzi}. 

In this letter, we present this general framework and exemplify it by constructing, for the first time, rotating black hole solutions of dimensionless spin up to 0.999 in scalar Gauss-Bonnet gravity, dynamical Chern-Simons gravity, and axi-dilaton gravity, all of which emerge as low-energy limits of string theory and quantum gravity candidates \cite{Maeda:2009uy, Moura:2006pz, Cano_Ruiperez_2019, Cano:2021rey, dCS_01, Taveras:2008yf}.
These new ``spectral black hole solutions'' show orders-of-magnitude improvement in spin in comparison with the previous series-in-$a$ solution, especially when considering rapidly-spinning black holes (see Fig.~2 and \cite{Lam:2025fzi}).
Moreover, the spectral solutions are analytic in nature, thus allowing us to investigate various astrophysical observables, e.g.~ringdown spectra, in modified gravity theories without being subjected to interpolation and other numerical error.

\vspace{0.2cm}
\noindent 
\textit{Effective field theory extension of general relativity.}\textemdash \
Consider a class of modified gravity theories that can be described by the Lagrangian density 
\begin{equation}
    \mathscr{L} = \frac{1}{16\pi} \Big[ R + \mathscr{L}_{\rm MG}[\varphi, g_{\mu\nu}] \Big], 
\end{equation}
where $R$ is the Ricci scalar associated with the metric $g_{\mu \nu}$ with signature $(-, +, +, +)$, $\varphi$ is a dynamical scalar or pseudoscalar field, $\mathscr{L}_{\rm MG}$ is the Lagrangian density modification to the Einstein-Hilbert action\footnote{In principle, dynamical vector fields can be included in $\mathscr{L}_{\rm MG}$, but here we focus on scalar-tensor modifications of gravity.}, and we henceforth employ units in which $G = 1 = c$. 
A general feature of modified gravity theories that are motivated by effective-field theory considerations is that $\mathscr{L}_{\rm MG}$ can be represented as a series in curvature invariants that may or may not be non-minimally coupled to dynamical scalar fields \cite{Maeda:2009uy, Moura:2006pz, Cano_Ruiperez_2019, Cano:2021rey, dCS_01, Taveras:2008yf}. 
One can then write $\mathscr{L}_{\rm MG} = \mathscr{L}_{\rm quad} + \mathscr{L}_{\rm cubic} + \ldots$, to represent quadratic and cubic curvature terms, such as \cite{Yunes:2011we, Cano_Ruiperez_2019, Yagi:2015oca, dCS_02, Bueno:2016xff} 
\begin{align}\label{eq:quadratic_Lagrangian}
    \mathscr{L}_{\rm quad} &= -\frac{1}{2} (\nabla \varphi)^2 - V(\varphi) + \alpha f(\varphi) \mathscr{Q}(R), \\
    \mathscr{L}_{\rm cubic} &= \lambda_{\rm ev} R_{\mu\nu}{}^{\rho\sigma} R_{\rho\sigma}{}^{\delta \gamma} R_{\delta \gamma}{}^{\mu\nu} + \lambda_{\rm odd} R_{\mu\nu}{}^{\rho\sigma} R_{\rho\sigma}{}^{\delta \gamma} \tilde{R}_{\delta \gamma}{}^{\mu\nu}, 
    \label{eq:cubic-Lagrangian}
\end{align}
where $V$ is a potential of the scalar field $\varphi$, $\alpha \sim \ell_{\rm quad}^2$ and $\lambda_{\rm ev, odd} \sim \ell_{\rm ev, odd}^4$ are dimensionful coupling constants of the theory with $\ell_{\rm quad, ev, odd}$ as their bare coupling constants, $f$ is a coupling function of the field $\varphi$, $\mathscr{Q}$ is a quadratic topological invariant, and ${R}_{\delta \gamma}{}^{\mu\nu}$ and $\tilde{R}_{\delta \gamma}{}^{\mu\nu}$ are the Riemann tensor and its dual. 
Since in many quantum gravity frameworks, shift-symmetry ($\varphi \sim \varphi + c$, $\forall c \in \mathbb{R}$) is manifest in their Lagrangian density, which implies that $V'(\varphi) = 0$ and $f(\varphi) = \varphi$. 
We will restrict our computations to such cases only. 

For a gravity theory that involves a massless scalar field (such as those governed by $\mathscr{L}_{\rm quad}$ with $V'(\varphi) = 0$ and $f(\varphi) = \varphi$ in Eq.~\eqref{eq:quadratic_Lagrangian}), we vary the action with respect to $\varphi$ and $g^{\mu\nu}$ to obtain the equation of motion of the scalar field and the modified field equations, namely 
\begin{align}
    \Box \varphi + \alpha \mathscr{Q}(R) &= 0, \label{eq:BoxAndFieldEquation1} \\
    G_{\mu\nu} - \mathscr{S}_{\mu\nu}(\varphi, g_{\mu\nu}) &= 0, 
    \label{eq:BoxAndFieldEquation2}
\end{align}
where $\Box = \nabla_{\mu}\nabla^{\mu}$ is the d’Alembertian operator, $G_{\mu\nu}$ is the Einstein tensor, and
\begin{equation}
\mathscr{S}_{\mu\nu} = -\frac{2}{\sqrt{-g}} \frac{\delta (\sqrt{-g}\mathscr{L}_{\rm MG})}{\delta g^{\mu\nu}}
\end{equation}
is the source tensor. 
The specific expression of $\mathscr{S}_{\mu\nu}$ depends on the modified theory of choice. If the modified theory does not include scalar fields, one only has to solve Eq.~(\ref{eq:BoxAndFieldEquation2}), with the scalar field set to zero, to obtain the leading order metric corrections.

Given that general relativity has passed all tests to date, we assume the modified theory introduces small deformations away from Einstein's theory, which amounts to requiring that the coupling be suitably small. 
For a black hole spacetime in the example-theories presented above, this ``small-coupling'' condition reads $\ell_{\rm quad, ev, odd} \ll M$, where $M$ is the (ADM) mass of the black hole. 
Then, the metric and scalar field can be expanded order-by-order in the coupling constant. 
To facilitate this ``small-coupling'' expansion, we define a dimensionless coupling constant $\zeta$, where $\zeta = \alpha^2/M^4$ in the case of $\mathscr{L}_{\rm quad}$, and $\zeta = \lambda_{\rm ev, odd}/M^4$ for $\mathscr{L}_{\rm cubic}$. 
The expansion can hence be written as 
\begin{equation}
    g_{\mu\nu} = g_{\mu\nu}^{(0)} + \zeta g_{\mu\nu}^{(1)} + O(\zeta^2), \quad 
    \varphi = \zeta^{1/2} \varphi^{(1/2)} + O(\zeta^{3/2}), 
\end{equation}
where $g_{\mu\nu}^{(0)}$ is the metric in general relativity, $\varphi^{(1/2)}$ is the rescaled background scalar field, and $g_{\mu\nu}^{(1)}$ is the leading-order-in-$\zeta$ metric correction. 
Expanding Eqs.~(\ref{eq:BoxAndFieldEquation1}) and~\eqref{eq:BoxAndFieldEquation2} perturbatively in $\zeta$, the leading-order equations are 
\begin{align}
    E_{\varphi} \equiv \Box^{(0)} \varphi^{(1/2)} + \zeta^{1/2} \mathscr{Q}^{(0)}(R^{(0)}) &= 0, \label{eq:SF0} \\
    E_{\mu\nu} \equiv G_{\mu\nu}^{(1)}(g_{\mu\nu}^{(0)}, g_{\mu\nu}^{(1)}) - \mathscr{S}_{\mu\nu}^{(0)}(\varphi^{(1/2)}, g_{\mu\nu}^{(0)}) &= 0. \label{eq:EE1}
\end{align}
Solving Eq.~(\ref{eq:SF0}), we obtain $\varphi^{(1/2)}$, and we can then compute $\mathscr{S}_{\mu\nu}^{(0)}$, which is (usually) quadratic in $\varphi$ and of order ${\cal{O}}(\zeta)$.
With  $\mathscr{S}_{\mu\nu}^{(0)}$ computed, we then solve Eq.~\eqref{eq:EE1} to obtain $g_{\mu\nu}^{(1)}$. If the theory does not contain a scalar field, we then simply solve Eq.~\eqref{eq:EE1} with $\varphi^{(1/2)} = 0$, where $\mathscr{S}_{\mu\nu}^{(0)}$ is proportional to $\lambda_{\rm ev,odd}$, which is also of ${\cal{O}}(\zeta)$.

\vspace{0.2cm}
\noindent 
\textit{Rotating black hole solutions beyond general relativity via spectral expansions.}\textemdash \
Rotating (uncharged) black holes in general relativity are described by the Kerr solution, and thus we must set $g_{\alpha\beta}^{(0)} = g_{\alpha\beta}^{\rm Kerr}$ in Boyer-Lindquist coordinates. 
We here further choose coordinates $x^{\alpha} = (t,r,\chi,\phi)$, with $\chi = \cos{\theta}$, such that the metric can be re-parameterized as~\cite{Cano_Ruiperez_2019}
\begin{equation}\label{eq:metric}
g_{\alpha\beta} = g_{\alpha\beta}^{\rm Kerr} (1 + \zeta h_{\alpha\beta})\,,
\end{equation}
where $h_{\alpha \beta}$ is a metric deformation, so that $g_{\alpha\beta}^{(1)} = g_{\alpha\beta}^{\rm Kerr} h_{\alpha \beta}$, where no Einstein summation is implied, e.g.~$g_{tt}^{(1)} = g_{tt}^{\rm Kerr} h_{tt}$.
To recover the Kerr metric asymptotically at spatial infinity, it is necessary to impose a set of boundary conditions. Expanding $h_{\alpha\beta}(r, \chi) = h_{\alpha\beta}^{(0)}(\chi) + r^{-1} h_{\alpha\beta}^{(1)}(\chi) + O(r^{-2})$, one finds that~\cite{Cano_Ruiperez_2019}
\begin{equation}\label{eq:BoundaryCondition}
    -2h_{t\phi}^{(0)} = h_{rr}^{(0)} = h_{\chi\chi}^{(0)} = h_{\phi\phi}^{(0)} = -\frac{h_{rr}^{(1)}}{M}, 
\end{equation}
and all other $h_{\alpha\beta}^{(0)} = 0$. 
With these conditions, the constants $M$ and $a$ that enter $g_{\alpha\beta}^{\rm Kerr}$ continue to represent the ADM mass and the dimensionless spin of the black hole (related to the ADM angular momentum via $J = M^2 a$). 
This metric ansatz and coordinate system has the advantage that the event horizon is located at its Kerr location, $r_+ = M + M (1 - a^2)^{1/2}$, at the cost of the radial coordinate not coinciding with the areal radius.

Let us now represent the metric deformation through spectral expansions. Assuming $h_{\alpha\beta}$ and its derivatives are finite and smooth outside of the event horizon, we begin by projecting $h_{\alpha\beta}(r, \chi)$ onto an orthogonal spectral basis $u_{n\ell}(r, \chi)$, 
\begin{align}\label{eq:Ansatz1}
    \varphi(r,\chi) &= \sum_{n, \ell = 0}^N c_{n\ell} u_{n\ell}(r, \chi)\,, \\
    h_{\alpha\beta}(r, \chi) &= \sum_{n, \ell = 0}^N v_{\alpha\beta,n\ell} u_{n\ell}(r, \chi), 
    \label{eq:Ansatz2}
\end{align}
where we truncate the expansion at spectral order $N$. 
We here choose our basis to be a product of Chebyshev polynomials in the compactified coordinate $z = {2r_+}/{r} - 1$ (so that the domain $r \in (r_+,\infty)$ is mapped to $z \in (1,-1)$), and Legendre polynomials in $\chi$, i.e., $u_{n\ell} = T_{n}(z)P_{\ell}(\chi)$.  
Spectral representations of this type are advantageous for solving coupled, linear, partial differential equations thanks to their exponential convergence with $N$ \cite{boyd2013chebyshev} and the ability to impose an ansatz with suitable symmetries and asymptotic properties.

This spectral expansion representation allows us to convert the partial-differential system of Eqs.~\eqref{eq:SF0} and~\eqref{eq:EE1} into a linear algebra problem for the coefficients $c_{n \ell}$ and $v_{\alpha \beta, n \ell}$. 
In the small coupling approximation, these equations are not coupled, but rather, they must be solved separately and in a given order (starting with the scalar field), as explained above. 
The linear algebra problem for the scalar field coefficients $c_{n \ell}$ has already been discussed in \cite{Stein:2014xba}, so we do not repeat it here, and instead postpone a detailed presentation to our companion paper \cite{Lam:2025fzi}. 
The linear algebra problem for the metric deformation coefficients $v_{\alpha \beta, n \ell}$ can be written as 
$\mathbb{D} \mathbf{v} = \mathbf{s}$, where $\mathbf{v}$ is a flattened list of $v_{\alpha\beta, n\ell}$, the matrix $\mathbb{D}$ is
\begin{equation}
    \mathbb{D}_{IJ} = \int u_{n'\ell'} (\mathscr{D}^{\alpha\beta})_{\mu\nu} u_{n\ell} \, d\mu_u, 
\end{equation}
where $\mathscr{D}^{\alpha\beta}{}_{\mu\nu}$ is a linear differential operator in $(r,\chi)$ that depends on $(M,a)$ and represents $G_{\mu\nu}^{(1)}$, while the vector $\mathbf{s}$ is 
\begin{equation}
    \mathbf{s}_J = \int u_{n'\ell'} \mathscr{S}_{\mu\nu}^{(0)} \, d\mu_u, 
\end{equation}
where $\mathscr{S}_{\mu\nu}^{(0)}$ is the source that appears in Eq.~\eqref{eq:EE1}. 
In these equations, $I = (\alpha,\beta,n,\ell)$ and $J = (\mu,\nu,n',\ell')$ are composite indices, and $d\mu_u$ is the integration measure with an appropriate weight function for $u_{n\ell}$. 
The matrix $\mathbb{D}_{IJ}$ and the vector $\mathbf{s}_J$ depend only on the properties of $u_{n\ell}$ and the precise form of the field equations, which are known analytically; therefore, they can be evaluated symbolically as functions of $(M,a)$ only.

To incorporate the constraints Eqs.~(\ref{eq:BoundaryCondition}) to the matrix equation, we project the constraint equations onto the angular part of $u_{n\ell}$, which transforms the constraint equations into a matrix constraint equation $\mathbb{A} \mathbf{v} = \mathbf{0}$.
We use these constraints to eliminate a subset of variables in $\mathbf{v}$, forming a reduced matrix equation $\tilde{\mathbb{D}} \tilde{\mathbf{v}} = \mathbf{s}$, where $\tilde{\mathbb{D}}$ is a rectangular matrix and $\tilde{\mathbf{v}}$ contains the remaining variables. 
Finally, the spectral solution is found by numerically inverting the matrix equation, $\tilde{\mathbf{v}} = (\tilde{\mathbb{D}}^{\rm T} \tilde{\mathbb{D}})^{-1} \tilde{\mathbb{D}}^{\rm T} \mathbf{s}$, using these $\tilde{\mathbf{v}}$ to solve for the remaining elements of ${\mathbf{v}}$ from the constraint equation, and then reconstructing $h_{\alpha\beta}$ from Eq.~(\ref{eq:Ansatz2}). 

Observe that the solution $h_{\alpha\beta}$ and $\varphi$ are explicit and analytic functions of the coordinates $(r,\chi)$, which are known in closed form. 
The only numerical ingredient in the solution is the numerical inversion used to obtain the spectral coefficients $\tilde{\mathbf{v}}$. 
Given numerical solutions for the spectral coefficients at various spins, one can construct a numerical interpolating function, or an analytic and closed-form representation. 
The latter can be found by fitting the numerical solutions for $v_{\alpha \beta, n \ell}$ and $c_{n \ell}$ at various spins to a given closed-form function of $a$ only. 
Doing so, one obtains a fully analytic and closed-form solution for the metric and scalar field of non-Kerr black holes (as a function of coordinates $(r,\chi)$ and spin $a$) in a wide class of series that is valid to leading-order in the small coupling approximation but to all orders in spin. 

\vspace{0.2cm}
\noindent 
\textit{Three examples.}\textemdash \
To exemplify our spectral framework and demonstrate its validity, we apply it to three specific cases of quadratic gravity: scalar Gauss-Bonnet gravity, dynamical Chern-Simons gravity, and axi-dilaton gravity. 
The Lagrangian density of these theories are that of Eq.~(\ref{eq:quadratic_Lagrangian}), with ${\cal{L}}_{\rm MG} = {\cal{L}}_{\rm quad}$, and with $V = 0$ (massless scalars). 
The coupling constants of these theories in ${\cal L}_{\rm MG}$ are observed to be small \cite{Chung:2025wbg,Lyu:2022gdr,Silva:2022srr,Silva:2020acr}, and thus, it is valid to apply this perturbative-in-$\zeta$ framework.
In scalar Gauss-Bonnet and in dynamical Chern-Simons gravity, $(\nabla \varphi)^2 = (\nabla \varphi_{\rm sGB,dCS})^2$,  $f(\varphi) = \varphi_{\rm sGB, dCS}$ (shift-symmetry), and $\mathscr{Q} = \mathscr{Q}_{\rm sGB,dCS}$, respectively, where $\mathscr{Q}_{\rm sGB} = R_{\mu\nu\rho\sigma} R^{\mu\nu\rho\sigma} - 4 R_{\mu\nu} R^{\mu\nu} + R^2$ is the Gauss-Bonnet invariant \cite{Yagi:2015oca, Nair:2019iur} and $\mathscr{Q}_{\rm dCS} = \tilde{R}^{\mu\nu\rho\sigma} R_{\mu\nu\rho\sigma}$ is the Pontryagin invariant~\cite{dCS_01, dCS_02}. 
In axi-dilaton gravity, $(\nabla \varphi)^2 = (\nabla \varphi_{\rm sGB})^2 + (\nabla \varphi_{\rm dCS})^2$ and $f(\varphi) \mathscr{Q} = \varphi_{\rm GB} \mathscr{Q}_{\rm sGB} + \varphi_{\rm dCS} \mathscr{Q}_{\rm dCS}$. 
These coupling functions $f(\varphi)$ can be thought of as a leading-order, small-field approximation of a more sophisticated $f(\varphi)$, when $\mathscr{Q}$ is a topological invariant \cite{Chung:2024vaf}. 
Given these Lagrangian densities, one can easily derive the field equations, and the source tensor $\mathscr{S}_{\mu\nu}$ can be computed explicitly (see e.g.~\cite{Cano_Ruiperez_2019,Yunes:2009hc}).

\begin{figure}[t!]
    \centering
    \includegraphics[width=\columnwidth]{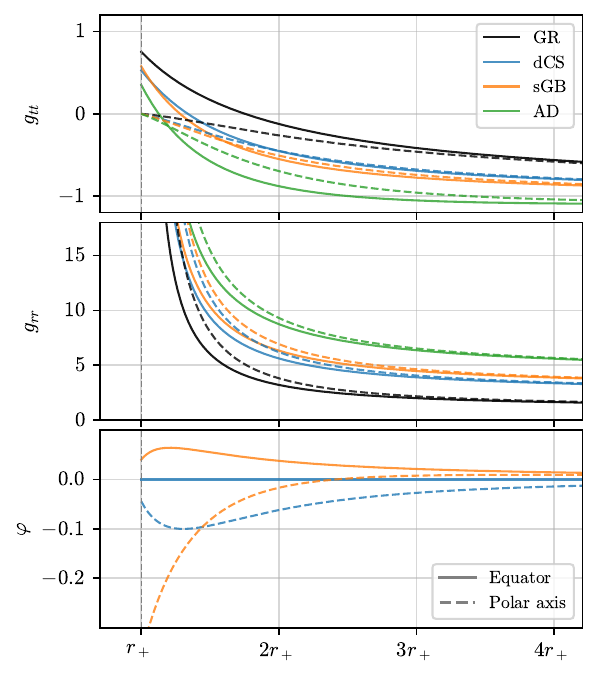}
    \caption{Scalar field and metric components on the equator (solid) and polar axis (dashed). 
    The top, middle and bottom panels display the behavior of the $g_{tt}$ and $g_{rr}$ components of the metric and the scalar field $\varphi$ respectively, all for a Kerr (black), a scalar Gauss-Bonnet (orange), a dynamical Chern-Simons (blue), and an axi-dilaton gravity (green) black hole with coupling constant $\zeta = 0.01$ and dimensionless spin $a = 0.99$. 
    Observe that the non-Kerr black holes have a smaller ergosphere and an event horizon at $r = r_+$, while their scalar fields have different parity depending on the theory considered. 
    }
    \label{fig:gtt}
\end{figure}

We can now implement the spectral expansion framework to find stationary and axisymmetric, black hole solutions of any non-extremal spin\footnote{Extremal black holes are subtle to handle, as they are parameterically very close to naked singularities; as such, we shall study these solutions elsewhere.} (for the metric deformation and the scalar field) in scalar Gauss-Bonnet, dynamical Chern-Simons, and axi-dilaton gravity.
Figure~\ref{fig:gtt} shows $g_{tt}$, $g_{rr}$, and $\varphi$ for Kerr and non-Kerr black holes with dimensionless spin $a = 0.99$ and coupling constant $\zeta = 0.01$, on the equator and polar axis, and as a function of radius, as a representative example of the full solution. 
The behavior of $g_{tt}$ on the equator and along the polar axis signals that these non-Kerr black holes have a smaller ergosphere (i.e.~their ergospheres are less ellipsoidal) than their Kerr counterparts, as is also the case for slowly-rotating black holes constructed from slow-rotation expansions \cite{Cano_Ruiperez_2019}. 
The behavior of $g_{rr}$ on the equator signals that the radial coordinate of the non-Kerr black holes is not the areal radius as anticipated, i.e.~$g_{rr} \to h_{rr}^{(0)} \neq 1$ as $r \to \infty$, unlike for Kerr black holes~\cite{Cano_Ruiperez_2019}. 
The divergent behavior of $g_{rr}$ for non-Kerr black holes as $r \to r_+$ signals the presence of an event horizon, just as in the Kerr case, as required by Eq.~(\ref{eq:metric}). 
The behavior of the scalar field signals that $\varphi_{\rm dCS}$ vanishes on the equator due to its odd parity, while $\varphi_{\rm sGB}$ is even in $\chi$ and non-vanishing on the equator due to its even parity. 

To access the accuracy of these spectral solutions, we compute 
\begin{align}\label{eq:EEResidual}
     \mathcal{E}_{\varphi} &= \left[ \int_{r_+}^{\infty} \!\!\! \int_{-1}^{+1} E_{\varphi}^{2} \sqrt{-g^{(0)}} \, dr d\chi \right]^{1/2},  \\
    \mathcal{E} &= \bigg[ \int_{r_+}^{\infty} \!\!\! \int_{-1}^{+1} E_{\mu\nu}E^{\mu\nu} \, \frac{\Delta^4}{r^8} (1 - \chi^2)^2 \sqrt{-g^{(0)}}\,drd\chi \bigg]^{1/2}, 
\end{align}
where $E_{\varphi}$ and $[E_{\mu \nu}]^{(1)}$ were defined in Eqs.~\eqref{eq:SF0} and~\eqref{eq:EE1}. 
A solution that satisfies $E_{\mu \nu} = 0$ would have ${\cal{E}}=0$ and it would be exact at leading order in the small coupling approximation.  
The quantity $\Delta^4 r^{-8} (1 - \chi^2)^2$ is a simple regularization factor that eliminates singularities at the event horizon and at the poles. 
For a fixed value of $a$, we find that $\mathcal{E}_{\varphi}$ and $\mathcal{E}$ decrease approximately exponentially as $N$ increases, as expected from the exponential convergence of spectral expansions; as $a$ increases, the rate of convergence decreases but it remains exponential\footnote{See the companion paper \cite{Lam:2025fzi} for details about convergence and precision of the spectral solutions.}.
Empirically, we find that with $N = 50$ for the scalar fields and $N = 45$ for the metric deformations, $(\mathcal{E}_{\varphi},\mathcal{E}) \lesssim (10^{-16},10^{-12})$ for $a \leq 0.99$ and $(\mathcal{E}_{\varphi},\mathcal{E}) \lesssim (10^{-12},10^{-8}$) for $a \leq 0.999$.

We now fit the numerical solutions for the spectral coefficients $c_{n \ell}$ (with $N=50$) and $v_{\alpha \beta, n \ell}$ (with $N=45$) at various spins to an analytic and closed-form fitting function of $a$.
Empirically, we find that the following fitting functions
\begin{align}
    c_{n \ell} &= \sum_{m=0}^{40} c_{n \ell,m} (\log \delta)^m\,, \quad
    v_{\alpha \beta, n \ell} = \sum_{m=0}^{30} \nu_{\alpha \beta, n \ell,m} (\log \delta)^m\,,
\end{align}
yield an excellent fit for $a \in [0, 0.99]$, with overall fitting residuals $|v_{\alpha\beta,n\ell}^{\rm fit} - v_{\alpha\beta,n\ell}^{\rm spectral}| \lesssim 10^{-15}$ \cite{Lam:2025fzi}. 
Using these logarithmic-in-$a$ representations for $(c_{n \ell},v_{\alpha \beta, n \ell})$ in Eqs.~\eqref{eq:Ansatz1} and~\eqref{eq:Ansatz2}, we obtain analytic and closed-form solutions in $(r,\chi;a)$ for black holes in scalar-Gauss-Bonnet, dynamical Chern-Simons, and axi-dilaton gravity that are valid to leading-order in the small-coupling approximation and any spin. 

\begin{figure}[t!]
    \centering
    \includegraphics[width=\columnwidth]{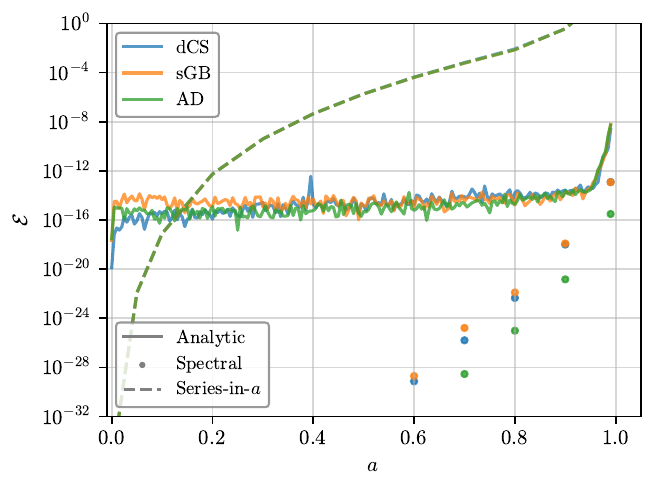}
    \caption{Error measure in the satisfaction of the field equations [Eq.~(\ref{eq:EEResidual})] when using the spectral (with $N=45$, markers), the analytic (solid), and the series-in-$a$ (to ${\cal{O}}(a^{15})$, dashed) black hole solutions in dynamical Chern-Simons (blue), scalar Gauss-Bonnet (orange), and axi-dilaton (green) gravity, as a function of black hole spin. 
    Note that the dashed lines for all three theories overlap completely. 
    Observe that the spectral and the analytic solutions are the only ones capable of accurately representing non-Kerr black hole solutions in these theories. 
    }
    \label{fig:MetricCorrectionsFit}
\end{figure}

To assess the accuracy of the spectral and analytic solutions across a wide range of spins, Fig.~\ref{fig:MetricCorrectionsFit} presents a measure of their error (in satisfying the field equations) ${\cal{E}}$, which we compare to the error obtained when using a series-in-$a$ solution (valid to ${\cal{O}}(a^{15})$) in the slow-rotation  \cite{Cano_Ruiperez_2019, Chung:2023zdq, Chung:2025gyg}. 
Observe that ${\cal{E}}$ is much smaller with the spectral solution than with the series-in-$a$ solutions, staying always below $10^{-12}$ for all spins. 
Observe also that ${\cal{E}}$ with the analytic solution is below $10^{-14}$ for spins $a < 0.99$, only increasing to $10^{-8}$ for near extremal black holes\footnote{Since the analytic solution is constructed from a fit of the spectral coefficients to the numerically-computed spectral coefficients over a wide range of spins (rather than e.g.~via interpolation), the fitted spectral coefficients do not reproduce exactly the numerically-computed coefficients.
Hence, the absolute error computed with the analytic solution differs from that computed with the spectral solution.
}.
The spectral and analytic solutions obtained here can be made even more accurate in a systematic way by increasing the number of basis functions ($N$), if this is ever needed.  
In particular, observe that the analytic solution becomes more accurate than the series-in-$a$ solution for $a > 0.15$.
Therefore, only the spectral and the analytic solutions are capable of accurately describing non-Kerr black holes in these theories when the spin is not small.

\vspace{0.2cm}
\noindent 
\textit{Future directions.}\textemdash \
The specific black hole solutions obtained here in scalar Gauss-Bonnet, dynamical Chern-Simons and axi-dilaton gravity can now be used to compute astrophysical observables associated with rapidly-rotating black holes, without incurring uncertainties due to finite-difference errors (especially from derivatives of the metric) or the slow-rotation expansion. 
Two key examples are the modeling of quasi-normal modes and extreme mass-ratio inspirals. In both cases, one can use the analytic solutions obtained here as a background on which to study linear vacuum perturbations (to obtain the quasi-normal spectrum, see e.g.~\cite{Chung:2024ira, Chung:2024vaf, Chung:2025gyg}) or matter perturbations (to obtain extreme mass-ratio inspiral waveforms). 
The former are particularly important to carry out ringdown tests of general relativity with gravitational waves detected by advanced LIGO and its partners, in a way that is robust to systematics, as described e.g.~in~\cite{Chung:2025wbg}. 
The latter are similarly important to carry out tests of general relativity with extreme-mass-ratio inspirals using LISA observations \cite{Gair:2017ynp, Chamberlain:2017fjl, Babak:2017tow, Maselli:2020zgv, Barsanti:2022vvl}. 

The framework presented here can also be applied to other beyond-Einstein theories to construct accurate, rapidly-rotating black hole solutions, such as in the cubic gravity theories of Eq.~\eqref{eq:cubic-Lagrangian}. 
These new solutions, together with the ones studied in this paper, are typically much more computationally efficient than slow-rotation expansions, requiring fewer terms (by a factor of at least two at moderate spins) to produce equally-accurate results. 
Moreover, these solutions allow for future, detailed investigations of the properties of nearly-extremal, non-Kerr black holes, including near-extremal observables, such as innermost stable circular orbits and photon  \cite{Lam:2025fzi}. 
The framework can also be applied to obtain spinning black hole solutions immersed in a non-vacuum environment \cite{Li:2025ffh}, such as around dark matter halos or accretion \cite{Fernandes:2025osu}. 
Therefore, our spectral expansions allow for the first accurate, computationally-efficient and analytic solutions for black holes of any spin in and outside general relativity.

\vspace{0.2cm}
\noindent 
\textit{Acknowledgments.}\textemdash \
The authors acknowledge the support from the Simons Foundation through Award No. 896696, the Simons Foundation International under grant SFI-MPS-BH-00012593-01, the NSF through Grant No. PHY-2207650 and PHY 25-12423, and NASA through Grant No. 80NSSC22K0806. 
A. K. W. C also acknowledge the Herchel Smith Fellowship at the University of Cambridge for partial support of this work. 
The calculations and results reported in this Letter were produced using the computational resources of the Illinois Campus Cluster, a computing resource that is operated by the Illinois Campus Cluster Program (ICCP) in conjunction with National Center for Supercomputing Applications (NCSA), and is supported by funds from the University of Illinois at Urbana-Champaign.
The author would like to specially thank the investors of the IlliniComputes initiatives and GravityTheory computational nodes for permitting the authors to execute runs related to this work using the relevant computational resources. 


\bibliography{ref.bib}

\end{document}